\documentclass{tlp}
\usepackage{latexsym}
\usepackage{amssymb}
\begin{document}
\submitted{16 October 2002}
\revised{7 August 2003}
\accepted{4 November 2003}
\title[Computing Convex Hulls with a Linear Solver]
{{\rm PROGRAMMING PEARL}\\[2ex]
Computing Convex Hulls with a Linear Solver}
\author[Florence Benoy, Andy King and Fred Mesnard]
{FLORENCE BENOY and ANDY KING\\
Computing Laboratory, University of Kent, UK. \\
email: \texttt{\{p.m.benoy, a.m.king\}@kent.ac.uk} \and
FRED MESNARD\\
Iremia, Universit\'e de La R\'eunion, France. \\
email: \texttt{fred@univ-reunion.fr}}
\maketitle

\newtheorem{lemma}{Lemma}[section]
\newtheorem{proposition}{Proposition}[section]
\newtheorem{theorem}{Theorem}[section]
\newtheorem{corollary}{Corollary}[section]
\newtheorem{definition}{Definition}[section]
\newtheorem{example}{Example}[section]

\def\real{\mathbb{Q}}
\def\rats{\mathbb{Q}}
\def\lbag{[\![}
\def\rbag{]\!]}
\def\PCH{P_{C\!H}}
\newcommand{\poly}[2]{{P_{#1, #2}}}

\begin{abstract}
A programming tactic involving polyhedra is reported that has been
widely applied in the polyhedral analysis of (constraint) logic
programs. The method enables the computations of convex hulls that
are required for polyhedral analysis to be coded with linear
constraint solving machinery that is available in many Prolog
systems.

\end{abstract}

\section{Introduction}

Polyhedra have been widely applied in program analysis
\cite{halbwachs} particularly for reasoning about logic and
constraint logic programs. In this context polyhedra have been
used in binding-time analysis \cite{bta}, cdr-coded list analysis
\cite{horspool-object}, argument-size analysis \cite{benoy},
time-complexity analysis \cite{lower}, high-precision groundness
analysis \cite{lpos}, type analysis \cite{saglam}, termination
checking \cite{basis} and termination inference
\cite{inference,lpar01}.

All these techniques use polyhedra to describe relevant properties
of the program and manipulate polyhedra using operations that
include projection, emptiness checking, inclusion testing for
polyhedra, intersection of polyhedra (meet) and the convex hull
(join). The classic approach to polyhedral analysis
\cite{halbwachs} uses two representations: (i) frames and rays and
(ii) systems of (non-strict) linear inequalities and employs the
Chernikova algorithm to convert between them \cite{leverge}. The
rationale for this dual representation is that the convex hull can
be computed straightforwardly with frames and rays whereas
intersection is more simply computed over systems of
linear inequalities. A simpler tactic that has been widely adopted
in the analysis of logic programs is to use only the linear
inequality representation and compute the convex hull by adapting
\cite{benoy} a relaxation technique proposed in \cite{BDB93}. The
elegance of this approach is that it enables the convex hull to be
computed without recourse to a dual representation: the problem is
recast as a projection problem that can be subcontracted to
standard linear constraint solving machinery with minimal coding
effort. Moreover, the performance is acceptable for many
applications. In fact this technique has been widely applied in
the analysis of logic programs
\cite{basis,lpar01,lower,inference,saglam}. The next section
outlines the method and the following section, an example
implementation. The final section presents the concluding
discussion.

\section{Method}\label{sect-method}

Consider two arbitrary polyhedra, $P_{1}$ and $P_{2}$,
represented in standard form:
\[
P_{1} = \{\vec{x} \in \rats^{n} \, | \, A_{1}\vec{x} \leq \vec{B}_{1}
\}
\qquad
P_{2} = \{\vec{x} \in \rats^{n} \, | \, A_{2}\vec{x} \leq \vec{B}_{2} \}
\]
such that $P_1 \neq \emptyset$ and $P_2 \neq \emptyset$  so that
the problem is non-trivial. Note that
$A_{i}\vec{x} \leq \vec{B}_{i}$ are non-strict and therefore
$P_1$ and $P_2$ are both closed. The problem in essence
is to compute the smallest
polyhedron that includes
$P_{1}$ and $P_{2}$. Interestingly, the convex
hull of $P_1 \cup P_2$ is not necessarily closed
as is illustrated in the following example.

\begin{example} \label{example-closure} \rm
Consider the 2-dimensional polyhedra $P_1$ and $P_2$ defined by:

\[
P_1 = \left\{ \vec{x} \in \real^2
\left|
\left[ \begin{array}{@{}rr@{}}
1 & 0 \\
-1 & 0 \\
0 & 1 \\
0 & -1 \\
\end{array}
\right]
\right.
\! \vec{x}
\leq
\left[ \begin{array}{@{}r@{}}
0 \\
0 \\
1 \\
-1 \\
\end{array}
\right]
\right\}
\quad
P_2 = \left\{ \vec{x} \in \real^2
\left|
\left[ \begin{array}{@{}rr@{}}
1 & -1 \\
-1 & 1 \\
-1 & 0 \\
\end{array}
\right]
\right.
\! \vec{x}
\leq
\left[ \begin{array}{@{}r@{}}
0 \\
0 \\
0 \\
\end{array}
\right]
\right\}
\]

\noindent Observe that
$P_{1} =
\left\{
\langle 0, 1 \rangle
\right\}$ is a point
whereas
$P_{2} =
\left\{
\langle x,y \rangle \in \real^{2}
\left| \,
x = y \wedge   0 \leq x
\right.
\right\}$ is a half-line.
Note too that $P_1$ and $P_2$ are closed
whereas
the convex hull
of $P_1 \cup P_2$ excludes the points
$\left\{
\langle x,y \rangle \in \real^{2}
\left| \,
x > 0 \wedge y = x + 1
\right.
\right\}$ and hence is not closed (see the diagram below).

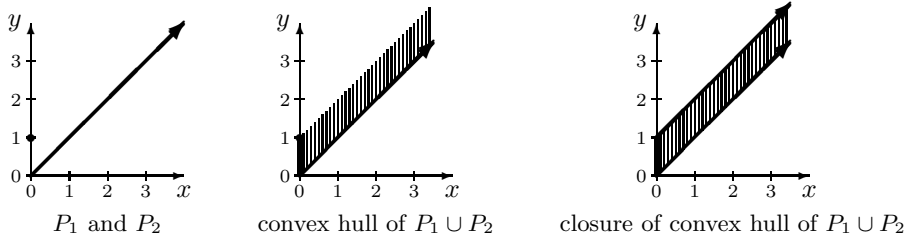
\begin{center}
\begin{tabular}{c@{\qquad}c@{\quad}c}
\setlength{\unitlength}{0.2in}
\begin{picture}(6,5.25)(-1,-1.25)
\thinlines

\put(0,0){\vector(1,0){4}}
\multiput(0,0.125)(1,0){4}{\line(0,-1){.25}}

\put(0,0){\vector(0,1){4}}
\multiput(-0.125,0)(0,1){4}{\line(1,0){.25}}

\put(0,-.25){\makebox(0,0)[t]{\scriptsize 0}}
\put(1,-.25){\makebox(0,0)[t]{\scriptsize 1}}
\put(2,-.25){\makebox(0,0)[t]{\scriptsize 2}}
\put(3,-.25){\makebox(0,0)[t]{\scriptsize 3}}
\put(4,-.25){\makebox(0,0)[t]{$x$}}

\put(-.25,0){\makebox(0,0)[r]{\scriptsize 0}}
\put(-.25,1){\makebox(0,0)[r]{\scriptsize 1}}
\put(-.25,2){\makebox(0,0)[r]{\scriptsize 2}}
\put(-.25,3){\makebox(0,0)[r]{\scriptsize 3}}
\put(-.25,4){\makebox(0,0)[r]{$y$}}

\put(0, 1){\circle*{0.2}}

\thicklines
\put(0,-0.03){\vector(1,1){4}}
\put(0,0.0){\vector(1,1){4}}
\put(0,0.03){\vector(1,1){4}}

\put(2,-1){\makebox(0,0)[t]{\small $P_1$ and $P_2$}}

\end{picture}

&

\setlength{\unitlength}{0.2in}
\begin{picture}(4,5.25)(0,-1.25)
\thinlines

\put(0,0){\vector(1,0){4}}
\multiput(0,0.125)(1,0){4}{\line(0,-1){.25}}

\put(0,0){\vector(0,1){4}}
\multiput(-0.125,0)(0,1){4}{\line(1,0){.25}}

\put(0,-.25){\makebox(0,0)[t]{\scriptsize 0}}
\put(1,-.25){\makebox(0,0)[t]{\scriptsize 1}}
\put(2,-.25){\makebox(0,0)[t]{\scriptsize 2}}
\put(3,-.25){\makebox(0,0)[t]{\scriptsize 3}}
\put(4,-.25){\makebox(0,0)[t]{$x$}}

\put(-.25,0){\makebox(0,0)[r]{\scriptsize 0}}
\put(-.25,1){\makebox(0,0)[r]{\scriptsize 1}}
\put(-.25,2){\makebox(0,0)[r]{\scriptsize 2}}
\put(-.25,3){\makebox(0,0)[r]{\scriptsize 3}}
\put(-.25,4){\makebox(0,0)[r]{$y$}}

\put(0, 1){\circle*{0.2}}

\thicklines \put(0,-0.03){\vector(1,1){3.5}}
\put(0,0.00){\vector(1,1){3.5}}
\put(0,0.03){\vector(1,1){3.5}}

\put(0.00,0.0){\line(0,1){1}}
\put(-0.03,0.0){\line(0,1){1}}
\put(0.03,0.0){\line(0,1){1}}

\thinlines
\multiput(0,0)(0.1,0.1){35}{\line(0,1){1}}

\put(2,-1){\makebox(0,0)[t]{\small convex hull of $P_1 \cup P_2$}}

\end{picture}

&

\setlength{\unitlength}{0.2in}
\begin{picture}(10,5.25)(-4,-1.25)
\thinlines

\put(0,0){\vector(1,0){4}}
\multiput(0,0.125)(1,0){4}{\line(0,-1){.25}}

\put(0,0){\vector(0,1){4}}
\multiput(-0.125,0)(0,1){4}{\line(1,0){.25}}

\put(0,-.25){\makebox(0,0)[t]{\scriptsize 0}}
\put(1,-.25){\makebox(0,0)[t]{\scriptsize 1}}
\put(2,-.25){\makebox(0,0)[t]{\scriptsize 2}}
\put(3,-.25){\makebox(0,0)[t]{\scriptsize 3}}
\put(4,-.25){\makebox(0,0)[t]{$x$}}

\put(-.25,0){\makebox(0,0)[r]{\scriptsize 0}}
\put(-.25,1){\makebox(0,0)[r]{\scriptsize 1}}
\put(-.25,2){\makebox(0,0)[r]{\scriptsize 2}}
\put(-.25,3){\makebox(0,0)[r]{\scriptsize 3}}
\put(-.25,4){\makebox(0,0)[r]{$y$}}

\thicklines \put(0,0.97){\vector(1,1){3.5}}
\put(0,-0.03){\vector(1,1){3.5}} \put(-0.03,0.0){\line(0,1){1}}

\put(0.0,1.00){\vector(1,1){3.5}}
\put(0.0,0.00){\vector(1,1){3.5}}
\put(0.00,0.0){\line(0,1){1}}

\put(0.0,1.03){\vector(1,1){3.5}}
\put(0.0,0.03){\vector(1,1){3.5}} \put(0.03,0.0){\line(0,1){1}}

\thinlines
\multiput(0,0)(0.1,0.1){35}{\line(0,1){1}}

\put(2,-1){\makebox(0,0)[t]{\small closure of convex hull of $P_1 \cup P_2$}}

\end{picture}

\end{tabular}
\end{center}
\end{example}

\noindent Since the convex hull of $P_1 \cup P_2$ is not
necessarily closed, the convex hull cannot always be represented
by a system of non-strict linear inequalities; in order
to overcome this problem,
the closure of the convex hull of $P_1 \cup P_2$ is computed.
The starting
point for our construction
is the convex hull of
$P_{1} \cup P_{2}$ that is given by:
\[
P_{H} = \left\{
\vec{x} \in \rats^{n} \,
\left|
\begin{array}{llllll}
\vec{x} = \sigma_{1}\vec{x}_{1} + \sigma_{2}\vec{x}_{2} & \wedge &
\sigma_{1} + \sigma_{2} = 1 & \wedge &  0 \leq \sigma_{1}  & \wedge \\
A_{1}\vec{x}_{1} \leq \vec{B}_{1} & \wedge &
A_{2}\vec{x}_{2} \leq \vec{B}_{2} & \wedge & 0 \leq \sigma_{2}
\end{array}
\right.
\right\}
\]

\noindent To avoid the non-linearity
$\vec{x} = \sigma_{1}\vec{x}_{1} + \sigma_{2}\vec{x}_{2}$
the system can be reformulated (relaxed) by putting
$\vec{y}_{1} = \sigma_{1}\vec{x}_{1}$ and
$\vec{y}_{2} = \sigma_{2}\vec{x}_{2}$ so that
$\vec{x} = \vec{y}_{1} + \vec{y}_{2}$ and
$A_{i}\vec{y}_{i} \leq \sigma_{i}\vec{B}_{i}$ to define:

\[
P_{C\!H} =
\left\{
\vec{x} \in \rats^{n} \,
\left|
\begin{array}{llllll}
\vec{x} = \vec{y}_{1} + \vec{y}_{2} & \wedge &
\sigma_{1} + \sigma_{2} = 1 & \wedge & 0 \leq \sigma_{1} & \wedge \\
A_{1}\vec{y}_{1} \leq \sigma_{1}\vec{B}_{1} & \wedge &
A_{2}\vec{y}_{2} \leq \sigma_{2}\vec{B}_{2} & \wedge & 0 \leq \sigma_{2}
\end{array}
\right.
\right\}
\]
Observe that $P_{H} \subseteq \PCH$.
Moreover, unlike $P_H$, $\PCH$ is expressed in terms of a system of
linear inequalities.
Note too that $\PCH$ is closed
since the projection of
a system of non-strict linear inequalities is closed.
In fact the following proposition
asserts that $\PCH$ coincides with the closure of the
convex hull of $P_1 \cup P_2$.

\begin{proposition} \label{prop}
$\PCH$ is the closure
of the convex hull of $P_1$ and $P_2$.
\end{proposition}

\noindent The proof uses the
concept of a recession cone.
The recession cone of a
polyhedron $P$, denoted
$0^{+}\!P$, is defined by:
$0^{+}P = \{ \vec{y} \in \rats^n \mid \forall
\lambda \geq 0 \; . \;
\forall \vec{x} \in P \; . \;
\vec{x} + \lambda \vec{y} \in P \}$.
The intuition is that $0^{+}P$ includes a vector $\vec{y}$
 whenever $P$
includes all the half-lines in the direction of $\vec{y}$ that
start in $P$.

\begin{proof}
Suppose $P_i = \{\vec{x} \in \rats^{n} \, | \, A_{i}\vec{x} \leq \vec{B}_{i} \}.$
Theorem 19.6 of \cite{rocky} states
that the closure of the convex hull of $P_1 \cup P_2$ is
the set
$({0^{+}}\!P_1 + P_2) \cup
(P_1 + {0^{+}}\!P_2) \cup
(\cup \{ \sigma_1 P_1 + \sigma_2 P_2 \mid \sigma_1 + \sigma_2 = 1
\wedge 0 \! < \sigma_1, \sigma_2 \}).$
Intuitively,
${0^{+}}\!P_1 + P_2$ is $P_2$ extended in the directions of half-lines
contained within $P_1$. Let
$\vec{x} \in P_i$, then
$\vec{y} \in {0^{+}}\!P_i$ if and only if
$A_{i}(\vec{x} + \lambda\vec{y}) \leq \vec{B_i}$ for all $\lambda \geq 0$ which holds
if and only if
$A_{i}\vec{y} \leq \vec{0}$ \cite{rocky}[pp 62].
Therefore
${0^{+}}\!P_1 + P_2 =
\{\vec{x} \in \rats^{n} \,| \,
\vec{x} = \vec{y}_{1} + \vec{y}_{2} \, \wedge \,
A_{1}\vec{y}_{1} \leq \vec{0}  \, \wedge \,
A_{2}\vec{y}_{2} \leq \vec{B}_{2}
\}$
and similarly
$P_1 \! + \! {0^{+}}\!P_2 =
\{\vec{x} \in \rats^{n} \, | \,
\vec{x} = \vec{y}_{1} + \vec{y}_{2} \, \wedge \,
A_{1}\vec{y}_{1} \leq \vec{B}_{1}  \, \wedge \,
A_{2}\vec{y}_{2} \leq \vec{0}
\}.$ Furthermore,
$\cup \{ \sigma_1 P_1 + \sigma_2 P_2 \mid \sigma_1 \! + \! \sigma_2 = 1
\, \wedge \, 0 < \sigma_1, \sigma_2 \}$ =
$\{\vec{x} \in \rats^{n} \mid
\sigma_{1} \! + \! \sigma_{2} = 1 \, \wedge \,
0 < \sigma_{1}, \sigma_{2} \, \wedge$
$\vec{x} = \vec{y}_{1} + \vec{y}_{2} \wedge \,
A_{1}\vec{y}_{1} \leq \sigma_{1}\vec{B}_{1}  \, \wedge \,
A_{2}\vec{y}_{2} \leq \sigma_{2}\vec{B}_{2}\}.$
Observe that
$\{\vec{x} \in \rats^{n} \mid
\vec{x} = \vec{y}_{1} + \vec{y}_{2} \wedge \,
A_{1}\vec{y}_{1} \leq \sigma_{1}\vec{B}_{1}  \, \wedge \,
A_{2}\vec{y}_{2} \leq \sigma_{2}\vec{B}_{2}\}$
coincides with
the sets
(i) ${0^{+}}\!P_1 + P_2$,
(ii) $P_1 \! + \! {0^{+}}\!P_2$
and
(iii) $\cup \{ \sigma_1 P_1 + \sigma_2 P_2 \mid \sigma_1 \! + \! \sigma_2 = 1
\, \wedge \, 0 < \sigma_1, \sigma_2 \}$
when
(i) $\sigma_{1} = 0$ and $\sigma_{2} = 1$,
(ii) $\sigma_{1} = 1$ and $\sigma_{2}= 0$
and
(iii) $\sigma_1 \! + \! \sigma_2 = 1$ and $0 < \sigma_1, \sigma_2$
respectively. Therefore
$\PCH$ is the closure of the convex hull.
\end{proof}

\noindent This result leads to an algorithm
for computing the closure of the convex hull: construct
the systems
$A_{i}\vec{y}_{i} \leq \sigma_{i}\vec{B}_{i}$ by scaling
the constant vectors $\vec{B}_{i}$ by $\sigma_{i}$,
add the constraints
$\vec{x} = \vec{y}_{1} + \vec{y}_{2}$,
$\sigma_{1} + \sigma_{2} = 1$ and $0 \leq \sigma_{i}$, then
eliminate variables other than $\vec{x}$ using projection to
obtain $\PCH$ in terms of $\vec{x}$.
Hence the closure of the convex hull can be computed without
recourse to another representation. This is illustrated below.

\begin{example} \label{abstract-example} \rm
Returning to example~\ref{example-closure},
consider the systems
$A_{i}\vec{x} \leq \vec{B}_{i}$:

\[
P_1 =
\left\{
\langle x, y \rangle \in \real^{2} \,
\left|
\begin{array}{@{\,}r@{\,}l@{\,}l@{\,}l@{}}
x \leq 0 & \wedge & -x \leq 0 & \wedge \\
y \leq 1 & \wedge & -y \leq -1 & \\
\end{array}
\right.
\right\}
\quad
P_2 =
\left\{
\langle x, y \rangle \in \real^{2} \,
\left|
\begin{array}{@{}r@{\,}l@{}}
x - y \leq 0 & \wedge \\
-x + y \leq 0 & \wedge \\
-x \leq 0 \\
\end{array}
\right.
\right\}
\]

\noindent Adding
$\vec{x} = \vec{y}_{1} + \vec{y}_{2},$
$\sigma_{1} + \sigma_{2} = 1$ and $0 \leq \sigma_{i}$ leads to the
following system:
\[
\PCH =
\left\{
\langle x, y \rangle \in \real^{2} \,
\left|
\begin{array}{llllll}
x = x_{1} + x_{2} & \wedge & y = y_{1} + y_{2} & \wedge &
\sigma_{1} + \sigma_{2} = 1 & \wedge  \\
0 \leq \sigma_{1} & \wedge & 0 \leq \sigma_{2} & \wedge \\
x_{1} \leq 0 & \wedge & -x_{1} \leq 0 & \wedge \\
y_{1} \leq \sigma_{1} & \wedge & -y_{1} \leq -\sigma_{1} & \wedge \\
x_{2} - y_{2} \leq 0 & \wedge & -x_{2} + y_{2} \leq 0 & \wedge &
-x_{2} \leq 0 \\
\end{array}
\right.
\right\}
\]
Eliminating the variables $x_{i},\,y_{i}$ and $\sigma_{i}$ leads
to the solution:
$$\PCH = \{\langle x, y \rangle \in \real^{2} \,| \,0 \leq x \wedge x \leq y \wedge y \leq x + 1\}$$
Theorem 19.6 of \cite{rocky}, which is used
in the proof,
asserts that $\PCH$ includes
$P_1 + {0^{+}}\!P_2  = P_1 + P_2 =
\{ \langle x, y \rangle \in \rats^2 \mid x \geq 0 \wedge y = x + 1 \}$
and therefore includes
the points $\{ \langle x, y \rangle \in \rats^2 \mid x > 0 \wedge y = x + 1 \}$,
and hence ensures closure.
Note that calculating $\PCH$ without the inequalities $0 \leq
\sigma_{1}$ and $0 \leq \sigma_{2}$ -- the relaxation advocated in
\cite{BDB93} for computing convex hull -- gives $ \{\langle x, y
\rangle \in \real^{2} \,| \,0 \leq x \}$ which is incorrect.
\end{example}

\section{Implementation}

This section shows how closure of the convex hull can be
implemented elegantly using a linear solver in particular the
CLP($\rats$) library \cite{Holzbaur95}. The behaviour of a
predicate is described with the aid of modes, that is, + indicates
an argument that should be instantiated to a non-variable term
when the predicate is called; - indicates an argument that should be
uninstantiated; and ? indicates an argument that may or may not be
instantiated \cite{iso}.

\subsection{Closed Polyhedra}
Closed polyhedra will be
represented by lists (conjunctions)  of linear constraints of the
form \mbox{$c ::= e \leq e$} $|$ \mbox{$e = e$} $|$ \mbox{$e \geq
e$} where expressions take the form \mbox{$e ::= x$} $|$ $n$ $|$
\mbox{$n * x$} $|$ \mbox{$-e$} $|$ \mbox{$e + e$} $|$
\mbox{$e - e$} and $n$ is a rational number and $x$ is a variable.
A convenient representation for a closed polyhedron
is a (non-ground) list of constraints. This representation is interpreted
with respect to a totally ordered (finite) set of variables.
The ordering
governs the mapping
of each variable to its specific dimension. In practise, the ordering
on variables is itself represented by the position of each variable
within a list. Specifically,
if $C$ is a list of linear
constraints $[c_1, \ldots, c_m]$ and $X$ is a list of variables
$[x_1, \ldots, x_n]$, then the represented polyhedron is
$\poly{C}{X} =
\{ \langle y_1, \ldots, y_n \rangle \in \real^n \mid
(\wedge_{i=1}^{n} x_i = y_i) \models_{\real} (\wedge_{j=1}^{m} c_j) \}$.
Note that although the order of variables
in $X$ is significant, the order of the constraints
in $C$ is not. Finally, let $vars(o)$ denote the set
of variables occurring in the syntactic object $o$.

\begin{example}\label{ex:p1p2}
The polyhedron $P_1$ from example \ref{abstract-example}
can be represented by the lists
$C_1 = [x=0$, \linebreak $y=1]$ and
$X = [x,y]$, that is,
$P_1 =  \poly{C_1}{X}$.
Moreover,
$P_2 = \poly{C_2}{X}$ where \linebreak
$C_2 = [x=y$, $x \geq 0]$ or alternatively
$C_2 = [y+z\geq x, x\geq y+2*z, y\geq 0, z \geq 0]$.
Hence the dimension of $\poly{C}{X}$ is defined
by the length of the list $X$ rather than the number of variables
in $C$.
\end{example}

\subsection{Projection}

Projection is central
to computing the convex hull. The desire, therefore, is
to construct a predicate 
{\tt project(+Xs,+Cxs,-ProjectCxs)} that is true when for
a given list of dimensions {\tt Xs} and a given list of constraints {\tt
Cxs}, {\tt ProjectCxs} is the projection of {\tt Cxs} onto
{\tt Xs}. The specification of such a predicate is given below.

\begin{description}
\item [preconditions:] \hspace{1cm}
  \begin{itemize}
    \item \texttt{Xs} is
          a closed list with distinct
          variables as elements,
    \item \texttt{Cxs} is a closed list of linear constraints,
    \item \texttt{Cxs} is satisfiable.
  \end{itemize}
\item [postconditions:] \hspace{1cm}
  \begin{itemize}
    \item \texttt{Xs} is
          a closed list with distinct
          variables as elements,
    \item \texttt{ProjectCxs} is a closed list of linear constraints,
 \item $vars(\mathtt{ProjectCxs}) \subseteq vars(\mathtt{Xs})$,
    \item $\poly{\mathtt{Cxs}}{\mathtt{Xs}} =
           \poly{\mathtt{ProjectCxs}}{\mathtt{Xs}}$.
  \end{itemize}
\end{description}

\noindent
Such a predicate can be constructed by
adding the given constraints to the store and then invoking
the projection facility
provided in the CLP($\rats$) library, that is,
the predicate
{\tt dump(+Target, -NewVars, -CodedAnswer)} \cite{Holzbaur95}.
Quoting from the manual: ``[dump] reflects the constraints
on the target variables into a term,
where {\tt Target} and {\tt NewVars} are lists
of variables of equal length and
{\tt CodedAnswer} is the term representation of the projection of constraints
onto the target variables where the target variables are
replaced by the corresponding variables from {\tt NewVars}''. 
This leads to the following implementation of {\tt project}:

\begin{verbatim}
    :- use_module(library(clpq)).

    project(Xs, Cxs, ProjectCxs) :-
        tell_cs(Cxs),
        dump(Xs, Vs, ProjectCxs), Xs = Vs.

    tell_cs([]).
    tell_cs([C|Cs]) :- {C}, tell_cs(Cs).
\end{verbatim}

\begin{example}
For example,
the query \texttt{project([X, Z], [X < Y, Y < Z], ProjectCs)} will
correctly bind \texttt{Cs} to \texttt{[X-Z<0]}.  However, correctness of
this predicate is compromised by existing
constraints in the store. 
For instance, the compound query
\texttt{\{X = Z + 1\}, project([X, Z], [X < Y, Y < Z], ProjectCs)}
will fail because constraints \linebreak posted
within {\tt tell\_cs} 
interact with those already in the store.
\end{example}

\noindent To insulate the constraints posted
in {\tt tell\_cs}, both
the variables {\tt Xs} and the constraints {\tt Cxs} need
to be renamed.
Renaming is trivial with the builtin \texttt{copy\_term}
but care must be taken to ensure that
{\tt Xs} and {\tt Cxs}
are renamed consistently, that is that variable
sharing in {\tt Xs} and {\tt Cxs} is preserved
in the copies.  However, in SICStus Prolog
{\tt copy\_term(Term, Cpy)}
copies any constraints in the store
that involve variables in {\tt Term}.
For example, the query
{\tt \{X=Y\}, copy\_term(X=Y+1, Cpy)}
will bind {\tt Cpy} to {\tt \_A=\_B+1}
where {\tt \_A} and {\tt \_B} are fresh variables. It will
also copy the constraint {\tt X = Y} by
posting the new constraint {\tt \_A = \_B} to the store.
To nullify this effect, {\tt copy\_term} is called within
the scope of {\tt call\_residue}. 
The call {\tt call\_residue(copy\_term(X=Y+1, Cpy), Residue)}
residuates any new constraint into {\tt Residue} instead of
posting it to the store, thereby copying the term without
copying any constraint. 
Whether residuation 
is required depends on the particular Prolog system.
This leads to the following (SICStus Prolog specific) revision:

\begin{verbatim}
    project(Xs, Cxs, ProjectCxs) :-
        call_residue(copy_term(Xs-Cxs, CpyXs-CpyCxs), _),
        tell_cs(CpyCxs),
        dump(CpyXs, Vs, ProjectCxs), Xs = Vs.
\end{verbatim}

\begin{example}
Using this revision, the query
\texttt{\{X = Z + 1\}, project([X, Z], [X < Y, Y < Z], ProjectCs)}
will succeed binding {\tt ProjectCs} to {\tt [X-Z<0]}.
However, adding {\tt Z = 5} to the list of constraints
induces an error. The problem is that posting the constraints
binds \texttt{Z} to 5
so that
\texttt{dump} is called with its first argument
instantiated
to a list that contains a non-variable term.
\end{example}

\noindent A pre-processing predicate
{\tt prepare\_dump} is therefore introduced to
ensure that {\tt dump} is called correctly.
The following revision to \texttt{project}, in effect,
extends the facility
provided by {\tt dump} to capture constraints over
both uninstantiated and instantiated variables:

\begin{verbatim}
    project(Xs, Cxs, ProjectCxs) :-
        call_residue(copy_term(Xs-Cxs, CpyXs-CpyCxs), _),
        tell_cs(CpyCxs),
        prepare_dump(CpyXs, Xs, Zs, DumpCxs, ProjectCxs),
        dump(Zs, Vs, DumpCxs), Xs = Vs.

    prepare_dump([], [], [], Cs, Cs).
    prepare_dump([X|Xs], YsIn, ZsOut, CsIn, CsOut) :-
        (ground(X) ->
            YsIn  = [Y|Ys],
            ZsOut = [_|Zs],
            CsOut = [Y=X|Cs]
        ;
            YsIn  = [_|Ys],
            ZsOut = [X|Zs],
            CsOut = Cs
        ),
        prepare_dump(Xs, Ys, Zs, CsIn, Cs).
\end{verbatim}

\noindent The literal \texttt{prepare\_dump(+Xs, +Ys, -Zs, ?CsIn, -CsOut)}
is true for a given list {\tt Xs} which contains either
variables or numbers (or a mixture of the two) and a given list
{\tt Ys} which contains only variables, if
\begin{itemize}

\item {\tt Zs} is the list obtained by substituting the non-variable
terms of {\tt Xs} with fresh variables and

\item {\tt CsOut} is an open ended list of equality constraints
with {\tt CsIn} at its end that contains one equality constraint
for each number in {\tt Xs}. Each constraint equates a numeric
element of {\tt Xs}
with the element of {\tt Ys} that is in the same list position.

\end{itemize}

\noindent The call \texttt{prepare\_dump([X1, 1, X3,
2], [A, B, C, D], Zs, CsIn, CsOut)}, for instance, will bind \texttt{Zs} to
\mbox{\texttt{[X1,\_A,X3,\_B]}} and \texttt{CsOut} to
\mbox{\texttt{[B=1,D=2|CsIn]}}. The predicate ensures
that {\tt dump} is called with its first argument bound to
a list of free variables even when the list {\tt Xs} includes numbers.
In the CLP($\real$)
library, numbers coincide with rationals which are represented as
compound (ground) terms of the form {\tt rat($n$, $d$)} where $n$
and $d$ are integers. The {\tt ground(X)} test effectively checks whether {\tt
X} is instantiated to a number; the test {\tt number(X)} is
inappropriate since it would always fail.

%

\begin{example} Consider again example \ref{ex:p1p2}.
The second representation of $P_2$ can be simplified by
using projection as follows:
\begin{verbatim}
| ?- Cs = [Y+Z>=X,X>=Y+2*Z,Y>=0,Z>=0], project([X,Y], Cs, ProjectCs).
ProjectCs = [Y>=0,X=Y] ? ;
no
\end{verbatim}
The system \texttt{Cs} is expressed over 3
variables and therefore
defines a 3 dimensional space.
Intuitively,
the projection onto {\tt [X, Y]} is the shadow cast
by $\poly{\mathtt{Cs}}{\mathtt{[X, Y, Z]}}$
onto the 2 dimensional space over {\tt X} and {\tt Y}.
The projection {\tt ProjectCs} in fact defines
a half-line confined to the first quadrant since,
by rearranging {\tt Cs}, it follows that
$\poly{\mathtt{Cs}}{\mathtt{[X, Y, Z]}} =
\{ \langle x, y, z \rangle \in \real^3 \mid x=y \wedge 0 \leq y \wedge z=0 \}$.
\end{example}

\subsection{Convex Hull}

The specification for the main predicate
\texttt{convex\_hull(+Xs, +Cxs, +Ys, +Cys, -Zs, -Czs)}, and then its
code, is given below.

\begin{description}
\item [preconditions:] \hspace{1cm}
  \begin{itemize}
    \item \texttt{Xs} is
          a closed list with distinct
          variables as elements and likewise for \texttt{Ys},
    \item \texttt{Xs} and \texttt{Ys} have the same length,
    \item $vars(\texttt{Xs}) \cap vars(\texttt{Ys}) = \emptyset$,
    \item \texttt{Cxs} and \texttt{Cys} are closed lists of linear constraints,
    \item \texttt{Cxs} and \texttt{Cys} are both satisfiable,
    \item $vars(\mathtt{Cxs}) \subseteq vars(\mathtt{Xs})$ and
    $vars(\mathtt{Cys}) \subseteq vars(\mathtt{Ys})$.
  \end{itemize}

\newpage
\item [postconditions:] \hspace{1cm}
  \begin{itemize}
    \item \texttt{Xs}, \texttt{Ys} and \texttt{Zs} are closed lists with
    distinct variables as elements,
    \item \texttt{Zs} is the same length as both \texttt{Xs} and \texttt{Ys},
    \item \texttt{Czs} is a closed list of linear constraints,
    \item $vars(\mathtt{Czs}) \subseteq vars(\mathtt{Zs})$ and
          $(vars(\mathtt{Xs}) \cup vars(\mathtt{Ys})) \cap vars(\mathtt{Zs}) = \emptyset$,
    \item $\poly{\mathtt{Czs}}{\mathtt{Zs}}$ is
the closure of the convex hull of
$\poly{\mathtt{Cxs}}{\mathtt{Xs}} \cup \poly{\mathtt{Cys}}{\mathtt{Ys}}$.

  \end{itemize}
\end{description}
\noindent

\begin{verbatim}
    convex_hull(Xs, Cxs, Ys, Cys, Zs, Czs) :-
        scale(Cxs, Sig1, [], C1s),
        scale(Cys, Sig2, C1s, C2s),
        add_vect(Xs, Ys, Zs, C2s, C3s),
        project(Zs, [Sig1 >= 0, Sig2 >= 0, Sig1+Sig2 = 1|C3s], Czs).

    scale([], _, Cs, Cs).
    scale([C1|C1s], Sig, C2s, C3s) :-
        C1 =.. [RelOp, A1, B1],
        C2 =.. [RelOp, A2, B2],
        mul_exp(A1, Sig, A2),
        mul_exp(B1, Sig, B2),
        scale(C1s, Sig, [C2|C2s], C3s).

    mul_exp(E1, Sigma, E2) :- once(mulexp(E1, Sigma, E2)).

    mulexp(  X,   _,     X) :- var(X).
    mulexp(N*X,   _,   N*X) :- ground(N), var(X).
    mulexp( -X, Sig,    -Y) :- mulexp(X, Sig, Y).
    mulexp(A+B, Sig,   C+D) :- mulexp(A, Sig, C), mulexp(B, Sig, D).
    mulexp(A-B, Sig,   C-D) :- mulexp(A, Sig, C), mulexp(B, Sig, D).
    mulexp(  N, Sig, N*Sig) :- ground(N).

    add_vect([], [], [], Cs, Cs).
    add_vect([U|Us], [V|Vs], [W|Ws], C1s, C2s) :-
        add_vect(Us, Vs, Ws, [W = U+V|C1s], C2s).
\end{verbatim}

The predicate {\tt mulexp(?E1, ?Sigma, -E2)} scales
the numeric constants that occur within {\tt E1} by
the variable {\tt Sigma}, providing
they are not coefficients of variables, to obtain
the expression {\tt E2}. Note that {\tt Sigma} is a variable and
the expression {\tt E1}
may be a variable,
hence both {\tt E1} and {\tt Sigma} have mode ? rather than +.
Since
a non-ground representation is employed for expressions, the
test {\tt var(X)} is used to determine whether the expression
is a variable.
As before, the test {\tt ground(N)} detects numeric
constants -- rational numbers -- which are the only type
of subexpressions that are ground. Observe
that {\tt mulexp} can return more than one
solution, for example,
{\tt mulexp(X, Sig, E2)} generates
\mbox{\tt E2 = X};
\mbox{\tt X = -(\_A), E2 = -(\_A)};
\mbox{\tt X = -(-(\_A))}, \mbox{\tt E2 = -(-(\_A))} etc as
solutions. Thus the pruning operator {\tt once}
is applied within {\tt mul\_exp(?E1, ?Sigma, -E2)}
to prevent erroneous solutions.

The predicate {\tt scale(+C1s, ?Sigma, ?C2s, -C3s)} scales
each constraint within the list {\tt C1s} by the
variable {\tt Sigma}.  Each constraint consists of
a binary operator and two expressions, and
scaling is applied to the numeric
constants in each expression as specified by {\tt mul\_exp}.
For example, {\tt scale([X+2 >= 1+Y, Y = Z], Sigma, Tail, ScaledCs)}
binds {\tt ScaledCs} to {\tt [Y = Z, X+2*Sigma >= 1*Sigma+Y | Tail]}.
Note that \texttt{scale} finesses the problem of putting
\texttt{Cxs} and \texttt{Cys} into the standard form
\mbox{$A_{i}\vec{y}_{i} \leq \vec{B}_{i}$} before applying scaling.
In standard form, \mbox{\texttt{X+2 >= 1+Y}} is
\mbox{\texttt{Y-X =< 1}} but scaling constants on both sides of
the relational operator preserves equivalence in that
\mbox{\texttt{X+2*Sig >= 1*Sig+Y}} is equivalent to
\mbox{\texttt{Y-X =< 1*Sig}}.  The use of a difference list
avoids an unnecessary call to append in the body of {\tt convex\_hull}.

The predicate {\tt add\_vect(+Us, +Vs, -Ws, ?C1s, -C2s)} operates
on the lists {\tt Us = [U$_1$, $\ldots$, U$_n$]} and {\tt Vs =
[V$_1$, $\ldots$, V$_n$]} which correspond to the vectors
$\vec{y}_1$ and $\vec{y}_2$ (as introduced in
section~\ref{sect-method}). The argument {\tt Ws} is instantiated
to another list of variables {\tt [W$_1$, $\ldots$, W$_n$]}, which
corresponds with $\vec{x}$. The predicate creates the system of
equalities {\tt [W$_1$ = U$_1$+V$_1$, $\ldots$, W$_n$ =
U$_n$+V$_n$]} corresponding to the system $\vec{x} = \vec{y}_1 +
\vec{y}_2$. The scaled constraints output by the two calls to {\tt
scale} are passed to {\tt add\_vect} via its accumulator and
thereby combined with the system of equalities. For example, the
call {\tt add\_vect([X1,Y1], [X2, Y2], Ws, Tail, Cs)} returns the
bindings {\tt Cs = [\_A=Y1+Y2,\_B=X1+X2|Tail]} and {\tt Ws =
[\_B,\_A]}.

The predicate
{\tt convex\_hull(Xs, Cxs, Ys, Cys, Zs, Czs)} takes, as
input, two lists of constraints  ({\tt Cxs} and {\tt Cys}) and their
corresponding lists of variables ({\tt Xs} and {\tt Ys}) and produces as output
a single list of constraints {\tt Czs} over the variables {\tt Zs}
that represents the closure of the convex hull of
the two input polyhedra.
If {\tt Xs} and {\tt Ys}
are not variable disjoint, then the pre-requisite
can be satisfied
by appropriately renaming variables. Specifically, the
variables {\tt Xs} and constraints {\tt Cxs} can be
renamed with
{\tt copy\_term(Xs-Cxs, CpyXs-CpyCxs)}
and the call
{\tt convex\_hull(Xs, Cxs, Ys, Cys, Zs, Czs)} replaced
with
{\tt convex\_hull(CpyXs, CpyCxs, Ys, Cys, Zs, Czs)}.
Since the integrity
of the constraint store is preserved by {\tt project} and
since {\tt project} is the only source of interaction with
the store, then it follows that \texttt{convex\_hull}
also does not side-effect any existing constraints. The following
is an illustrative example.

\begin{example}
Running this code on the data of Example \ref{abstract-example}
gives: 
\begin{verbatim}
| ?- convex_hull([X1,Y1],[X1=0,Y1=1],[X2,Y2],[X2>=0,Y2=X2],V,S).
S = [_A>=0,_A-_B>=-1,_A-_B=<0],
V = [_A,_B] ? ;
no
\end{verbatim}
%
\end{example}


\setlength{\unitlength}{0.13in}
\begin{figure}[t]
\begin{center}
\begin{tabular}{c@{\qquad\qquad}c}

\begin{picture}(12,12)(-6,-6)
\thinlines
\put(-1,-1){\makebox(0,0)[t]{$P_1$}}
\put(1,1){\makebox(0,0)[b]{$P_2$}}

\put(-6,0){\vector(1,0){12}}
\multiput(-6,-0.125)(2,0){6}{\line(0,1){.25}}
\put(-4,-0.25){\makebox(0,0)[t]{\scriptsize -1}}
\put(4,-0.25){\makebox(0,0)[t]{\scriptsize 1}}
\put(6,-0.25){\makebox(0,0)[t]{$x$}}

\put(0,-6){\vector(0,1){12}}
\multiput(-0.125,-6)(0,2){6}{\line(1,0){.25}}
\put(-.25,-4){\makebox(0,0)[r]{\scriptsize -1}}
\put(-.25,4){\makebox(0,0)[r]{\scriptsize 1}}
\put(-.25,6){\makebox(0,0)[r]{$y$}}

\thicklines

\put(0,4){\line(1,-1){4}}
\put(0,-4){\line(-1,1){4}}

\put(-4,0){\line(1,0){8}}
\put(0,4){\line(0,-1){8}}

\end{picture}

&

\begin{picture}(12,12)(-6,-6)
\thinlines

\put(-1,-1){\makebox(0,0)[t]{$Q_1$}}
\put(1,1){\makebox(0,0)[b]{$Q_2$}}

\put(-6,-3){\vector(2,1){12}}
\multiput(-6,-2.875)(2,1){6}{\line(0,-1){.25}}
\put(-4,-2.25){\makebox(0,0)[t]{\scriptsize -1}}
\put(4,1.75){\makebox(0,0)[t]{\scriptsize 1}}
\put(6,2.75){\makebox(0,0)[t]{$x$}}

\put(0,-6){\vector(0,1){12}}
\multiput(-0.125,-6)(0,2){6}{\line(1,0){.25}}
\put(-.25,-4){\makebox(0,0)[r]{\scriptsize -1}}
\put(-.25,4){\makebox(0,0)[r]{\scriptsize 1}}
\put(-.25,6){\makebox(0,0)[r]{$y$}}

\put(-6,3){\vector(2,-1){12}}
\multiput(-6,2.875)(2,-1){6}{\line(0,1){.25}}
\put(-4,1.75){\makebox(0,0)[t]{\scriptsize -1}}
\put(4,-2.25){\makebox(0,0)[t]{\scriptsize 1}}
\put(6,-3.25){\makebox(0,0)[t]{$z$}}

\thicklines

\put(0,4){\line(2,-1){4}}

\put(0,4){\line(2,-3){4}}

\put(0,-4){\line(-2,1){4}}

\put(-4,2){\line(0,-1){4}}
\put(4,2){\line(0,-1){4}}

\put(0,0){\line(0,1){4}}
\put(0,0){\line(2,-1){4}}
\put(0,0){\line(-2,1){4}}
\put(0,0){\line(0,-1){4}}

\put(0,0){\line(-2,-1){4}}

\multiput(0,0)(0.2,0.1){20}{\circle*{0.01}}

\thinlines

\multiput(-4,2)(0.1,-0.15){40}{\circle*{0.01}}

\end{picture}

\\

\begin{picture}(12,12)(-6,-6)
\thinlines

\put(-6,0){\vector(1,0){12}}
\multiput(-6,-0.125)(2,0){6}{\line(0,1){.25}}
\put(-4,-0.25){\makebox(0,0)[t]{\scriptsize -1}}
\put(4,-0.25){\makebox(0,0)[t]{\scriptsize 1}}
\put(6,-0.25){\makebox(0,0)[t]{$x$}}

\put(0,-6){\vector(0,1){12}}
\multiput(-0.125,-6)(0,2){6}{\line(1,0){.25}}
\put(-.25,-4){\makebox(0,0)[r]{\scriptsize -1}}
\put(-.25,4){\makebox(0,0)[r]{\scriptsize 1}}
\put(-.25,6){\makebox(0,0)[r]{$y$}}

\thicklines

\put(0,4){\line(1,-1){4}}
\put(0,4){\line(-1,-1){4}}
\put(0,-4){\line(1,1){4}}
\put(0,-4){\line(-1,1){4}}

\end{picture}

&

\begin{picture}(12,12)(-6,-6)
\thinlines

\put(-6,-3){\vector(2,1){12}}
\multiput(-6,-2.875)(2,1){6}{\line(0,-1){.25}}
\put(-4,-2.25){\makebox(0,0)[t]{\scriptsize -1}}
\put(4,1.75){\makebox(0,0)[t]{\scriptsize 1}}
\put(6,2.75){\makebox(0,0)[t]{$x$}}

\put(0,-6){\vector(0,1){12}}
\multiput(-0.125,-6)(0,2){6}{\line(1,0){.25}}
\put(-.25,-4){\makebox(0,0)[r]{\scriptsize -1}}
\put(-.25,4){\makebox(0,0)[r]{\scriptsize 1}}
\put(-.25,6){\makebox(0,0)[r]{$y$}}

\put(-6,3){\vector(2,-1){12}}
\multiput(-6,2.875)(2,-1){6}{\line(0,1){.25}}
\put(-4,1.75){\makebox(0,0)[t]{\scriptsize -1}}
\put(4,-2.25){\makebox(0,0)[t]{\scriptsize 1}}
\put(6,-3.25){\makebox(0,0)[t]{$z$}}

\thicklines

\put(0,4){\line(2,-1){4}}
\put(0,4){\line(-2,-1){4}}

\put(0,4){\line(2,-3){4}}
\put(0,4){\line(-2,-3){4}}

\put(0,-4){\line(2,1){4}}
\put(0,-4){\line(-2,1){4}}

\put(-4,2){\line(0,-1){4}}
\put(4,2){\line(0,-1){4}}

\put(-4,-2){\line(1,0){8}}

\thinlines

\multiput(-4,2)(0.2,0){40}{\circle*{0.01}}
\multiput(-4,2)(0.1,-0.15){40}{\circle*{0.01}}
\multiput(4,2)(-0.1,-0.15){40}{\circle*{0.01}}

\end{picture}
\end{tabular}
\end{center}
\caption{
(i) $P_1$ and $P_2$,
(ii) $Q_1$ and $Q_2$,
(iii) $conv(P_1 \cup P_2)$,
(iv) $conv(Q_1 \cup Q_2)$}\label{figure-hello}
\end{figure}
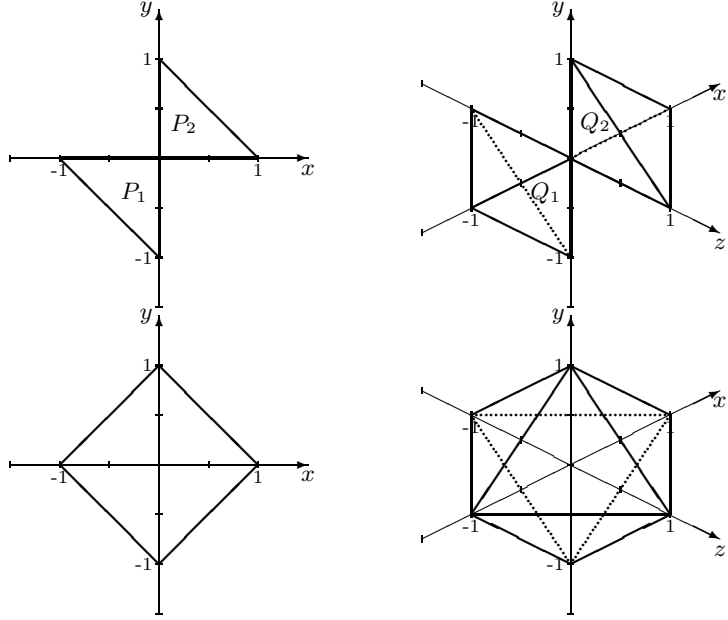

\section{Discussion}

This section discusses
the method proposed in the paper, comparing it
with related techniques.
The Chernikova method
is exponential in the worst-case \cite{leverge} and
the Fourier-Motzkin method, like all projection
techniques over linear inequalities \cite{chandru}, is also exponential.
The exponential behaviour of
both methods stems from the same source: the possibly
exponential relationship
between the number of vertices and the number of
half-spaces that define a polyhedron.  In fact
the problem of calculating the closure of
the convex hull of two polyhedra is also
exponential even for bounded polyhedra (polytopes).
This can be demonstrated by considering the so-called cross polytope
in $n$-dimensions which is the polyhedron with
the vertex set
$\{ \langle {\pm}1, 0, \ldots, 0 \rangle,
\langle 0, {\pm}1, \ldots, 0 \rangle,
\ldots,
\langle 0, 0, \ldots, {\pm}1 \rangle
\}$. The cross polytope can be defined by no less than $2^n$ inequalities
yet can arise as the convex hull of two polyhedra both
of which can be defined with $O(n)$ inequalities. Specifically consider
the $n$-dimensional polyhedra
\[
\begin{array}{r@{\;}l}
P_1 = & \{
\langle x_1, \ldots, x_n \rangle \in \real^n \mid
(\sum_{i=1}^{n} -{x_i}  \leq 1) \wedge (\wedge_{j=1}^{n} x_j \leq 0)
\}
\\
P_2 = & \{
\langle x_1, \ldots, x_n \rangle \in \real^n \mid
(\sum_{i=1}^{n} x_i  \leq 1) \wedge (\wedge_{j=1}^{n} -{x_j} \leq 0)
\}
\end{array}
\]
Because $P_1$ and $P_2$ are polytopes, they can be expressed in terms
of their vertices:
\[
\begin{array}{r@{\;}r@{\;}r@{\;}r@{\;}l}
P_1 = conv(\{
\langle 0, 0, \ldots, 0 \rangle, &
\langle -1, 0, \ldots, 0 \rangle, &
\langle 0, -1, \ldots, 0 \rangle, &
\ldots, &
\langle 0, 0, \ldots, -1 \rangle \})
\\
P_2 = conv(\{
\langle 0, 0, \ldots, 0 \rangle, &
\langle 1, 0, \ldots, 0 \rangle, &
\langle 0, 1, \ldots, 0 \rangle, &
\ldots, &
\langle 0, 0, \ldots, 1 \rangle \})
\end{array}
\]
Since $\langle 0, 0, \ldots, 0 \rangle$ is convexly spanned by
$\langle 1, 0, \ldots, 0 \rangle$
and
$\langle -1, 0, \ldots, 0 \rangle$, it follows that
$cl(conv(P_1 \cup P_2)) = conv(P_1 \cup P_2) =
conv(\{\langle {\pm}1, 0, \ldots, 0 \rangle$,
$\langle 0, {\pm}1, \ldots, 0 \rangle$,
$\ldots$,
$\langle 0, 0, \ldots, {\pm}1 \rangle \})$ which is the $n$-dimensional
cross polytope. The 2 and 3 dimensional cases are denoted
in Figure~\ref{figure-hello} by
(i) $P_1$ and $P_2$
and
(ii) $Q_1$ and $Q_2$ respectively
for which the cross polytopes are
a solid square and an octahedron. Hence the
problem of calculating the closure of the convex hull
is intrinsically
exponential irrespective of the algorithm employed.

\begin{example}  The following query illustrates
how the hull algorithm yields an exponential number of
inequalities for the 4 dimensional case.
\begin{verbatim}
| ?- Xs = [X1, X2, X3, X4], Ys = [Y1, Y2, Y3, Y4],
     Cxs = [-1 =< X1+X2+X3+X4, X1 =< 0, X2 =< 0, X3 =< 0, X4 =< 0],
     Cys = [ Y1+Y2+Y3+Y4 =< 1, 0 =< Y1, 0 =< Y2, 0 =< Y3, 0 =< Y4],
     convex_hull(Xs, Cxs, Ys, Cys, Zs, Czs),
     Zs = [A, B, C, D].

Czs = [A-B+C+D>=-1, A+B-C-D=<1, A+B+C+D>=-1, A-B-C-D=<1,
       A-B-C+D>=-1, A+B+C-D=<1, A+B-C+D>=-1, A-B+C-D=<1,
       A-B+C-D>=-1, A+B-C+D=<1, A+B+C-D>=-1, A-B-C+D=<1,
       A-B-C-D>=-1, A+B+C+D=<1, A+B-C-D>=-1, A-B+C+D=<1] ? ;

no
\end{verbatim}
\end{example}

\noindent However, it would be wrong to conclude
from these examples that the frame and ray representation
is preferable --  inequalities are unavoidable since
they are required for other polyhedral operations.

Despite the scaling problems that are inherent
to any convex hull algorithm, in practise the
technique proposed in this paper has been
widely applied in logic programming
\cite{basis,lpar01,lower,inference,saglam}, mostly to satisfaction.
For example, in the context of inferring termination
conditions for logic programs
this method is feasible since
it accounts for
 42\% of this first pass of the analysis and
the first pass
itself constitutes only 23\% of the total analysis time \cite{inference}.
Whether the approach presented in
this paper is applicable depends on the application context.
When only standard domain operations
are required and performance
is not an issue, this method has much to commend it.
However, when the application has to additionally
reason, say, about integral points \cite{ancourt,zpolyhedra} or
parameterised polyhedra \cite{loechner}
then specialised polyhedral libraries are required.
Further, if performance
is important, then recourse should be made to a polyhedral
library, since a state-of-the-art implementation employing
the Chernikova algorithm \cite{parma},
will outperform the approach presented here.

We have presented a Prolog program for computing convex hulls
using linear solver machinery. As Holzbaur's library is also
available for CIAO Prolog, ECLiPSe, XSB and Yap
Prolog, the technique can be easily adapted to these systems. The
method is a reasonable compromise between conciseness, clarity
and efficiency and variants of this program have now been
widely deployed.

\paragraph{Acknowledgements}
Thanks are due to
\mbox{Mats Carlsson}, 
\mbox{Bart Demoen},
\mbox{Pat Hill},
\linebreak \mbox{Joachim Schimpf} and
\mbox{Raimund Seidel}
and, of course, the anonymous referees.


\begin{thebibliography}{}

\bibitem[\protect\citeauthoryear{Ancourt}{Ancourt}{1991}]{ancourt}
{\sc Ancourt, C.} 1991.
\newblock G\'{e}n'{e}ration {A}utomatique de {C}ode de {T}ransfert pour
  {M}ultiprocesseurs \`{a} {M}\'{e}moires {L}ocales.
\newblock Ph.D. thesis, Universit\'{e} Paris 6.

\bibitem[\protect\citeauthoryear{Bagnara, Ricci, Zaffanella, and Hill}{Bagnara
  et~al\mbox{.}}{2002}]{parma}
{\sc Bagnara, R.}, {\sc Ricci, E.}, {\sc Zaffanella, E.}, {\sc and} {\sc Hill,
  P.~M.} 2002.
\newblock Possibly {N}ot {C}losed {C}onvex {P}olyhedra and the {P}arma
  {P}olyhedra {L}ibrary.
\newblock In {\em Static {A}nalysis {S}ymposium}, {M.~V. Hermenegildo} {and}
  {G.~Puebla}, Eds. Lecture Notes in Computer Science, vol. 2477.
  Springer-Verlag, 213--229.
\newblock See also http://www.cs.unipr.it/ppl/.

\bibitem[\protect\citeauthoryear{Benoy and King}{Benoy and King}{1996}]{benoy}
{\sc Benoy, F.} {\sc and} {\sc King, A.} 1996.
\newblock Inferring {A}rgument {S}ize {R}elationships with {CLP(${\cal R}$)}.
\newblock In {\em Logic-based {P}rogram {S}ynthesis and {T}ransformation
  ({S}elected {P}apers)}, {J.~P. Gallagher}, Ed. Lecture Notes in Computer
  Science, vol. 1207. Springer-Verlag, 204--223.

\bibitem[\protect\citeauthoryear{Chandru, Lassez, and Lassez}{Chandru
  et~al\mbox{.}}{2000}]{chandru}
{\sc Chandru, V.}, {\sc Lassez, C.}, {\sc and} {\sc Lassez, J.-L.} 2000.
\newblock Qualitative {T}heorem {P}roving in {L}inear {C}onstraints.
\newblock In {\em International {S}ymposium on {A}rtificial {I}ntelligence and
  {M}athematics}.
\newblock Long version to appear in the {A}nnals of {M}athematics and
  {A}rtificial {I}ntelligence.

\bibitem[\protect\citeauthoryear{Codish, Genaim, Sondergaard, and
  Stuckey}{Codish et~al\mbox{.}}{2001}]{lpos}
{\sc Codish, M.}, {\sc Genaim, S.}, {\sc Sondergaard, H.}, {\sc and} {\sc
  Stuckey, P.} 2001.
\newblock Higher-{P}recision {G}roundness {A}nalysis.
\newblock In {\em International {C}onference on {L}ogic {P}rogramming},
  {P.~Codognet}, Ed. Lecture Notes in Computer Science, vol. 2237.
  Springer-Verlag, 135--149.

\bibitem[\protect\citeauthoryear{Codish and Taboch}{Codish and
  Taboch}{1999}]{basis}
{\sc Codish, M.} {\sc and} {\sc Taboch, C.} 1999.
\newblock A {S}emantic {B}asis for the {T}ermination {A}nalysis of {L}ogic
  {P}rograms.
\newblock {\em The {J}ournal of {L}ogic {P}rogramming\/}~{\em 41,\/}~1,
  103--123.

\bibitem[\protect\citeauthoryear{Cousot and Halbwachs}{Cousot and
  Halbwachs}{1978}]{halbwachs}
{\sc Cousot, P.} {\sc and} {\sc Halbwachs, N.} 1978.
\newblock Automatic {D}iscovery of {L}inear {R}estraints among {V}ariables of a
  {P}rogram.
\newblock In {\em Principles of {P}rogramming {L}anguages}. ACM Press, 84--97.

\bibitem[\protect\citeauthoryear{{De Backer} and Beringer}{{De Backer} and
  Beringer}{1993}]{BDB93}
{\sc {De Backer}, B.} {\sc and} {\sc Beringer, H.} 1993.
\newblock A {CLP} language handling disjunctions of linear constraints.
\newblock In {\em International {C}onference on {L}ogic {P}rogramming}, {D.~S.
  Warren}, Ed. MIT Press, 550--563.

\bibitem[\protect\citeauthoryear{Deransart, Ed-Dbali, and Cervoni}{Deransart
  et~al\mbox{.}}{1996}]{iso}
{\sc Deransart, P.}, {\sc Ed-Dbali, A.}, {\sc and} {\sc Cervoni, L.} 1996.
\newblock {\em Prolog: {T}he {S}tandard}.
\newblock Springer-Verlag.

\bibitem[\protect\citeauthoryear{Genaim and Codish}{Genaim and
  Codish}{2001}]{lpar01}
{\sc Genaim, S.} {\sc and} {\sc Codish, M.} 2001.
\newblock Inferring {T}ermination {C}onditions for {L}ogic {P}rograms using
  {B}ackwards {A}nalysis.
\newblock In {\em International {C}onference on {L}ogic for {P}rogramming,
  {A}rtificial {I}ntelligence and {R}easoning}, {R.~Nieuwenhuis} {and}
  {A.~Voronkov}, Eds. Lecture Notes in Artificial Intelligence, vol. 2250.
  Springer-Verlag, 681--690.

\bibitem[\protect\citeauthoryear{Holzbaur}{Holzbaur}{1995}]{Holzbaur95}
{\sc Holzbaur, C.} 1995.
\newblock {OFAI} {clp(Q,R)} {M}anual.
\newblock Tech. Rep. TR-95-09, Austrian Research Institute for Artificial
  Intelligence (\"{O}FAI), Schottengasse 3, A-1010 Vienna, Austria.

\bibitem[\protect\citeauthoryear{Horspool}{Horspool}{1990}]{horspool-object}
{\sc Horspool, N.} 1990.
\newblock Mode {A}nalysis {T}echniques for {D}iscovery of {L}ists in {P}rolog.
\newblock In {\em Object {M}anagement}, {D.~Tsichritzis}, Ed. Centre
  Universitaire d'Informatique, University of Geneva, 305--312.

\bibitem[\protect\citeauthoryear{King, Shen, and Benoy}{King
  et~al\mbox{.}}{1997}]{lower}
{\sc King, A.}, {\sc Shen, K.}, {\sc and} {\sc Benoy, F.} 1997.
\newblock Lower-bound {T}ime-{C}omplexity {A}nalysis of {L}ogic {P}rograms.
\newblock In {\em International {S}ymposium on {L}ogic {P}rogramming},
  {J.~Maluszynski}, Ed. MIT Press, 261--276.

\bibitem[\protect\citeauthoryear{{Le Verge}}{{Le Verge}}{1992}]{leverge}
{\sc {Le Verge}, H.} 1992.
\newblock A note on {C}hernikova's algorithm.
\newblock Tech. Rep. 635, IRISA, Campus Universitaire de Beaulieu, Rennes,
  France.

\bibitem[\protect\citeauthoryear{Loechner and Wilde}{Loechner and
  Wilde}{1997}]{loechner}
{\sc Loechner, V.} {\sc and} {\sc Wilde, D.~K.} 1997.
\newblock Parameterized {P}olyhedra and their {V}ertices.
\newblock {\em International {J}ournal of {P}arallel {P}rogramming\/}~{\em
  25,\/}~6, 525--549.
\newblock See also http://icps.u-strasbg.fr/polylib/.

\bibitem[\protect\citeauthoryear{Mesnard and Neumerkel}{Mesnard and
  Neumerkel}{2001}]{inference}
{\sc Mesnard, F.} {\sc and} {\sc Neumerkel, U.} 2001.
\newblock Applying {S}tatic {A}nalysis {T}echniques for {I}nferring
  {T}ermination {C}onditions of {L}ogic {P}rograms.
\newblock In {\em Static {A}nalysis {S}ymposium}, {P.~Cousot}, Ed. Lecture
  Notes in Computer Science, vol. 2126. Springer-Verlag, 93--110.

\bibitem[\protect\citeauthoryear{Quinton, Rajopadhye, and Risset}{Quinton
  et~al\mbox{.}}{1997}]{zpolyhedra}
{\sc Quinton, P.}, {\sc Rajopadhye, S.~V.}, {\sc and} {\sc Risset, T.} 1997.
\newblock On {M}anipulating {Z}-{P}olyhedra using a {C}anonical
  {R}epresentation.
\newblock {\em Parallel {P}rocessing {L}etters\/}~{\em 7,\/}~2, 181--194.

\bibitem[\protect\citeauthoryear{Rockafellar}{Rockafellar}{1970}]{rocky}
{\sc Rockafellar, R.~T.} 1970.
\newblock {\em Convex {A}nalysis}.
\newblock Princeton {U}niversity {P}ress.

\bibitem[\protect\citeauthoryear{Sa\u{g}lam and Gallagher}{Sa\u{g}lam and
  Gallagher}{1997}]{saglam}
{\sc Sa\u{g}lam, H.} {\sc and} {\sc Gallagher, J.~P.} 1997.
\newblock Constrained {R}egular {A}pproximation of {L}ogic {P}rograms.
\newblock In {\em Logic {P}rogramming {S}ynthesis and {T}ransformation
  ({S}elected {P}apers)}, {N.~E. Fuchs}, Ed. Springer-Verlag, 282--299.

\bibitem[\protect\citeauthoryear{Vanhoof and Bruynooghe}{Vanhoof and
  Bruynooghe}{2001}]{bta}
{\sc Vanhoof, W.} {\sc and} {\sc Bruynooghe, M.} 2001.
\newblock Binding-{T}ime {A}nnotations {W}ithout {B}inding-{T}ime {A}nalysis.
\newblock In {\em International {C}onference on {L}ogic for {P}rogramming,
  {A}rtificial {I}ntelligence and {R}easoning}, {R.~Nieuwenhuis} {and}
  {A.~Voronkov}, Eds. Lecture Notes in Artificial Intelligence, vol. 2250.
  Springer-Verlag, 707--722.

\end{thebibliography}

\end{document}